\documentclass[12pt]{iopart}
\usepackage{iopams}
\newcommand{\beao}{\begin{eqnarray*}}
\newcommand{\eeao}{\end{eqnarray*}}
\newcommand{\be}{\begin{equation}}\newcommand{\ee}{\end{equation}}
\newcommand{\bea}{\begin{eqnarray}}
\newcommand{\eea}{\end{eqnarray}}
\newcommand{\beq}{\begin{eqnarray}}
\newcommand{\eeq}{\end{eqnarray}}
\newcommand{\nn}{\nonumber}
\newcommand{\pa}{\partial}
\newcommand{\ep}{\epsilon}
\newcommand{\om}{\omega}

\renewcommand{\tr}{{\rm tr}}
\newcommand{\Ref}[1]{(\ref{#1})}
\begin{document}
\title[Casimir force beyond PFA]{Casimir force for a sphere in front of a plane beyond Proximity Force Approximation}

\author{M.  Bordag$^1$ and V.  Nikolaev$^2$}

\address{$^1$ Institute for Theoretical Physics, Leipzig University,
Vor dem Hospitaltore 1, D-04103 Leipzig}
\address{$^2$  Halmstad University, Box 823, S-30118 Halmstad}
\eads{\mailto{bordag@itp.uni-leipzig.de}}
\eads{\mailto{Vladimir.Nikolaev@ide.hh.se}}
\begin{abstract}
For the configuration of a sphere in front of a plane  we calculate the first two terms of the asymptotic expansion for small separation
of the Casimir force. We consider both Dirichlet and Neumann boundary conditions.
\end{abstract}
\section{Introduction and discussion}
High precision measurements of the Casimir force are of remarkable
actual interest. For instance, these measurements provide stronger
constraints on hypothetical interactions in the sub-micrometer
region \cite{Mostepanenkothisproceedings} and give hope to measure
thermal forces. The high precision is achieved in a setup of
measuring the force between a flat plate and a sphere. The typical
radius of the sphere is $R\approx 200\mu m$ (although   recently \cite{Decca2007} a sphere with $R\approx 80\mu m$ was used) and the separation
ranges from contact till $L\approx 1\mu m$. This setup provides a
large area of opposing surfaces avoiding difficulties from keeping
the surfaces in parallel. The ratio of these two sizes,
\be \ep=\frac{L}{R}\sim 10^{-3}
\label{ep}\ee
is a small parameter describing the deviation of the geometry from the plane parallel one.

The calculation of the Casimir force for arbitrarily shaped surfaces was a long standing problem and one was left with unsatisfactory approximate methods. The most important of them is the proximity force approximation (PFA) known from \cite{derjaguin}. The method takes the force density known from the plane parallel case and integrates it over the curved surface. Clearly, this methods works only for small deviations and from the very setup it was impossible to get higher order corrections or information on the precision of the approximation. Nevertheless, PFA worked remarkably well  in the comparison of the precision measurements \cite{Mohideen2006,Decca2007} with theoretical expectations. Along with this, from increasing the precision of the measurements there appears a call to go beyond \cite{Decca2007}.

The evaluation of the Casimir force is in general plagued by the  ultraviolet divergences even what concerns the force between distinct bodies although these are finite.
The point is that divergences are present in intermediate steps. One has to introduce some regularization, say in a mode sum representation, to subtract the divergent terms (even if these do not depend on the separation),  and to perform the limit of removing the regularization. This procedure works well in simple geometries with separating variables in the corresponding wave equation. However, for a more general geometry, where one would have to calculate eigenvalues numerically, this hampers the calculations quite much. 
 
There exists approximate methods which gave important results. For instance,  the semiclassical approximation in \cite{SP1,SP2} has put the PFA on a solid base within quantum field theory. Also the optical path approximation \cite{Jaffe:2003mb} was very important in
this context, since it also confirmed the leading-order PFA from a
field-theory perspective and gave somewhat better results at
next-to-leading order.
 An exceptional role play the so called 'world line methods' \cite{Gies:2003cv}, where a functional integral representation of the Greens function is computed numerically. Regrettably, this method is restricted to Dirichlet boundary conditions.

Progress happened recently with the representation of the Casimir interaction energy between two bodies in terms of a functional determinant which is free of ultraviolet divergences \cite{WIRZBA2006,EMIG2006}. For any fixed distance between the bodies, the energy is represented by multiple convergent sums and integrals. These allow for direct numerical evaluation \cite{WIRZBA2006}. However, for small $\ep$ like the value in \Ref{ep}, the computations become too complex and are restricted to $\ep \gtrsim 0.1$.

A solution of this problem was found in \cite{Bordag:2006vc} where the first correction beyond PFA was calculated for a cylinder in front of a plane. Technically, an asymptotic expansion for small $\ep$ of the representation in terms of a functional determinant was  derived. In the leading order, the PFA was re-obtained and in the  next-to-leading order the first correction beyond PFA was calculated analytically. This was done for Dirichlet and for Neumann boundary conditions,
\bea\label{cyl}
E^{\rm cyl}_{\rm  Dirichlet} &=&
-\frac{1}{L^2}\sqrt{\frac{R}{L}}\frac{\pi^3}{1920\sqrt{2}}
\left\{
1
+   { \frac{7}{36} }\frac{L}{R}
+O\left(\left(\frac{L}{R}\right)^2\right)
\right\} \, ,
 \\ \nn
E^{\rm cyl}_{\rm  Neumann} &=&
-\frac{1}{L^2}\sqrt{\frac{R}{L}}\frac{\pi^3}{1920\sqrt{2}}
\left\{
1
+   {\left( \frac{7}{36} -\frac{40}{3\pi^2}\right)}\frac{L}{R}
+O\left(\left(\frac{L}{R}\right)^2\right)
\right\} .
\eea
The sum of these two,  by virtue of the wave guide geometry, delivers at once   the result for the electromagnetic case. It is remarkable, that the result for the Dirichlet boundary conditions was with good precision confirmed by the world line methods in \cite{Gies:2006cq}.

In the present paper we use the method developed in \cite{Bordag:2006vc} to calculate the Casimir interaction energy for a sphere in front of a plane. We perform the calculation for Dirichlet and for Neumann boundary conditions. The result is
\bea\label{sph}
E^{\rm sphere}_{ \mbox{\small Dirichlet}}&=&
 -\frac{\pi^3}{1440} \frac{R}{L^2}
 \left\{
1
+ \  \frac{1}{3}  \ \frac{L}{R}
+O\left(\left(\frac{L}{R}\right)^2\right)
\right\} ,
\\ \nn
E^{\rm sphere}_{ \mbox{\small Neumann}}&=&
 -\frac{\pi^3}{1440} \frac{R}{L^2}
 \left\{
1
+ \ \left(  \frac{1}{3} -\frac{10}{\pi^2}\right)  \ \frac{L}{R}
+O\left(\left(\frac{L}{R}\right)^2\right)
\right\}\, .\eea
With these formulas we re-confirm the PFA in leading order. In next-to-leading order we
obtained  analytically the first correction beyond PFA. Again, as in the cylindrical case, the Dirichlet case is confirmed by the 'world line methods' calculation  in \cite{Gies:2006cq}.
Since in this geometry the polarizations of the electromagnetic field do not separate, \Ref{sph} does not include the result for the electromagnetic case which is left as a future work. A general approach for that was considered in  \cite{Emig:2007cf} along with specific calculations of the Casimir interaction between
spheres over a wide range of separations, including rather
short distances.
 However, if expecting the corresponding result to be of the same order as \Ref{sph} one can conclude that these corrections do not affect the current high precision measurements of the Casimir force but must be taken into account in future ones.

In the remaining part of this paper we follow \cite{Bordag:2006vc} and represent the calculations resulting in \Ref{sph}. The essentially new element is a certain asymptotic expansion of the Clebsch-Gordan coefficients which was not found in literature.

\section{Asymptotic expansion of the Casimir energy for small separation}

 The expression of the Casimir energy for a sphere in front of a plane in terms of a convergent functional determinant is known from \cite{WIRZBA2006,EMIG2006,Bordag:2006vc}. Here we use  the notations adopted in \cite{Bordag:2006vc}.  The energy under consideration is given by
\be\label{E1}E=\frac{1}{2\pi} \int_0^\infty d\om  \ \tr _{l,m} \ln\left(\delta_{l,l'}-A_{l,l'} \right).
\ee
Here the trace is over the orbital momenta $l$ and $m$ in a representation in orbital momentum basis of the matrix elements
\bea\label{Alls}A_{l,l'}&=&\sqrt{\frac{\pi}{2}}\sqrt{\frac{(2l+1)(2l'+1)}{2a\om}}
(-1)^{l+l'}\frac{I_{l'+\frac12}(\om R)}{K_{l+\frac12}(\om R)}
 \\ &&\nn
\cdot\sum_{l''=|l-l'|}^{l+l'}   (-i)^{l''}(2l''+1)K_{l''+\frac12}(2a\om)
\left(\begin{array}{ccc}l''&l&l'\\0&0&0\end{array}\right)
\left(\begin{array}{ccc}l''&l&l'\\0&-m&m\end{array}\right),
\eea
where $I_\nu$ and $K_\nu$ are modified Bessel functions, $a=L+R$ is the distance from the plane to the center of the sphere and the symbols in brackets are the 3j-symbols. It is convenient for the following to expand the logarithm in \Ref{E1},
\be\label{E2}E=\frac{1}{2\pi} \int_0^\infty d\om  \ \sum_{s=0}^\infty \frac{-1}{s+1} \sum_{m=-\infty}^\infty
\sum_{l=|m|}^\infty \sum_{l_1=|m|}^\infty \dots \sum_{l_s=|m|}^\infty
A_{l,l_1} A_{l_1,l_2}\dots A_{l_s,l}  \, .
\ee
For fixed $R$ and $a$, the integral and all sums in this formula are convergent. With $\ep=L/R$, \Ref{ep}, we note $a=R(1+\ep)$ and for small $\ep$ the convergence gets lost. As for the frequency $\om$ this can be seen from the arguments of the Bessel functions. Using their asymptotic behavior for large argument, $I_\nu(z)\sim exp(z)$ and $K_\nu(z)\sim exp(-z)$, the integrand in the matrix
element behaves like $exp(-2\om\ep)$ and ceases to decrease for $\ep=0$. Similar arguments hold for the sums involved. As a consequence, a simple expansion for small $\ep$ does not work. 

Next we rewrite Eq.\Ref{E2} by making the substitution $\om\to\om/R$ and by changing the summation indices in the products of the matrix elements,
%
%
%
\bea\label{E3}E&=&
-\frac{1}{2\pi R}
\sum_{s=0}^\infty \frac{1}{s+1}
\int_0^\infty d\om  \
\sum_{m=-\infty}^\infty
\sum_{l=|m|}^\infty \sum_{l_1=-l}^\infty \dots \sum_{l_s=-l}^\infty
\nn \\ &&
 \cdot A_{l,l+l_1} A_{l+l_1,l+l_2}\dots A_{l+l_s,l}  \,.
\eea

The strategy of the calculation for small $\ep$ is as follows. We assume that the dominating contribution comes from the large values of the frequency $\om$ and all summation indexes involved in \Ref{E3}. We are interested in the asymptotic expansion for small $\ep$, hence we substitute all sums by integrations. 
The error introduced in this way is supposed to be exponentially small like in the difference between sum and integral in the Abel-Plana formula. 
After that we make the substitution of variables in \Ref{E3},
\bea
&& \om=\frac{t}{\ep}\,\sqrt{1-\tau^2}, \quad l=\frac{t\tau}{\ep}\,,
\quad m=\sqrt{\frac{t}{\ep}}\,\tau\mu
\nn \\
&& l_1=\sqrt{\frac{4t}{\ep}}\,n_1, \dots , l_s=\sqrt{\frac{4t}{\ep}}\,n_s\,,
\label{subst}
\eea
 manifesting that the main contributions come from $\om\sim 1/\ep$, from the 'main diagonal'  matrix index  $l\sim1/\ep$, from the 'off-diagonal directions' $l_i\sim1/\sqrt{\ep}$ $(i=1,\dots,s)$ and from the magnetic quantum number   $ m\sim 1/\sqrt{\ep}$.   The variable $\tau$ is the cosine of the angle in the $(\om,l)$-plane. After this we expand the matrix elements for $\ep\to0$
and calculate the remaining integrals.

In the new variables the expression for the energy reads
\bea\label{E4}E&\simeq&
-\frac{\ep^{-2}}{4\pi R}
\sum_{s=0}^\infty \frac{1}{s+1}
\int_0^\infty dt \, t  \
\int_{0}^1 \frac{d\tau\,\tau}{\sqrt{1-\tau^2}}\,
\int_{-\infty}^\infty \frac{d\mu}{\sqrt{\pi}}\,
\nn \\ &&  ~~~~~~~~~~~~~~\times
\int_{-\infty}^\infty \frac{dn_1}{\sqrt{\pi}}\,
\dots
\int_{-\infty}^\infty \frac{dn_s}{\sqrt{\pi}}\
{\cal M}  \
\eea
where
\be
{\cal M}=\left(\frac{4\pi t}{\ep}\right)^{\frac{s+1}{2}}
A_{l,l+l_1}\cdot \dots \cdot A_{l+l_s,l}
\label{A2}\ee
collects the information from the matrix elements.

In the appendix A we calculate the asymptotic expansion of the matrix elements \Ref{Alls} for $\ep\to0$ with the variables substituted according to Eq. \Ref{subst}. Using  Eq.\Ref{Alls3} obtained there we get  for the asymptotic expansion of the energy the expression
\bea
E&\simeq&-\frac{\ep^{-2}}{4\pi R}
\sum_{s=0}^\infty \frac{1}{s+1}
\int_0^\infty dt \, t  \ e^{-2(s+1)t}
\int_{0}^1 \frac{d\tau\,\tau}{\sqrt{1-\tau^2}}\,
\int_{-\infty}^\infty \frac{d\mu}{\sqrt{\pi}}\ e^{-(s+1)\mu^2}
\nn \\ &&  ~~~~~~~~~~~~~~\times
\int_{-\infty}^\infty \frac{dn_1}{\sqrt{\pi}}\,
\dots
\int_{-\infty}^\infty \frac{dn_s}{\sqrt{\pi}}\
e^{-\eta_1}
{\cal M}_{as}
\label{E5}\eea
with
\be\label{eta1}\eta_1= \sum_{i=0}^{s}\left(n_i-n_{i+1}\right)^2.
\ee
Here,  we have  formally to put $n_0=n_{s+1}=0$. The non-leading information is contained in
\bea\label{M}{\cal M}^{\rm as}&=&1+\sqrt{\ep}\ \sum_{i=0}^s  a_{n_i,{n}_{i+1}}^{(1/2)}
   +\ep\left(\sum_{0\le i <  j\le s}  a_{n_i,{n}_{i+1}}^{(1/2)} a_{n_j,{n}_{j+1}}^{(1/2)}
 +\sum_{i=0}^s  a_{n_i,{n}_{i+1}}^{(1)}
 \right)+\dots\nn,
\eea
where the sums result from the products of the matrix elements in \Ref{A2}.
Now it is possible to carry out the integrations. This can be done in complete analogy to the cylindrical case. For instance,  in the leading order we note
\be\label{}\int_{-\infty}^\infty \frac{dn_1}{\sqrt{\pi}}\dots  \int_{-\infty}^\infty \frac{dn_{s} }{\sqrt{\pi}} \ \ e^{-\eta_1} \ = \frac{1}{\sqrt{s+1}}\,.
\ee
Another square root of $(s+1)$ comes from the integration over $\mu$ and two more inverse powers of $(s+1)$ come from the integration over $t$. Collecting all factors together we immediately  come to the  expression know from the PFA thus reconfirming that approximation independently. For the next-to-leading orders we also use the same techniques as in the cylindrical case. Thus the integrations over the $n_i$  are Gaussian integrations with a somehow involved combinatorics. These are completely described in the appendix C in \cite{Bordag:2006vc}\footnote{Regrettably, that appendix contains a number of misprints. These are corrected in the  electronic preprint, arXiv:hep-th/0602295v2.}. After that the integrations over the remaining variables can be carried out too. All these integrations are either Gaussian or simple exponentials with polynomial factors in front. However, the number of terms in the intermediate steps is quite large. In fact, the calculations were performed machined using to a large extend the same scripts which served in the cylindrical case.
The result is the remarkably simple coefficient $1/3$ in Eq.\Ref{sph}.

Finally, we describe the changes which come in if using Neumann boundary conditions instead of Dirichlet ones.
First of all, from the Neumann condition on the plane we have a sign change in the expression for the energy which now reads
\be\label{EN}E=\frac{1}{2\pi} \int_0^\infty d\om  \ \tr _{l,m} \ln\left(\delta_{l,l'}+A^N_{l,l'} \right).
\ee
The derivatives on the sphere result in changed matrix elements $A^N_{l,l'}$.
From the derivation of the matrix elements it follows that the radial arguments of the matrix elements $r$ and $r'$ are just in the arguments of the Bessel functions $I_{l'+\frac12}(\om r')$ and $K_{l+\frac12}(\om r)$ in \Ref{Alls} before putting them on the sphere, i.e., before putting $r=R$ and $r'=R$. The derivative in the Neumann boundary conditions acts just on these arguments. Consequently, we have to substitute $I_{l'+\frac12}(\om r')\to \left(r\frac{\pa}{\pa r}+u\right) I_{l'+\frac12}(\om r')$ where $u$ is the parameter which appears in Robin boundary conditions. In fact, it is possible to formulate different types of Neumann conditions by multiplying the wave function with some power of the radius, which by means of $r\pa_r r^u \Phi=r^u(r\pa_r+u)\Phi$ is equivalent to a Robin condition. For a single  sphere,  one has to put $u=1/2$ in the electromagnetic case.

Now we discuss the changes which happen in the asymptotic expansions of these two Bessel function in the matrix element. It can be easily verified that for the asymptotic expansion of the Bessel functions the following formulas hold,
\bea
I_u\equiv\left(r\pa_r+u\right)I_\nu(\nu z)&=&\nu z I'_\nu(\nu z)+u I_\nu(\nu z),
\\ \nn
K_u\equiv\left(r\pa_r+u\right)K_\nu(\nu z)&=&\nu z K'_\nu(\nu z)+u K_\nu(\nu z).
\eea
Using the asymptotic expansions these expressions can be rewritten in the form
\bea
I_u&=&\nu\sqrt{1+z^2}\,I_\nu(\nu z)
\left(\frac{u}{\nu\sqrt{1+z^2}}
+\frac{\sum\limits_{k\ge0}\frac{v_k}{\nu^k}}{\sum\limits_{k\ge0}
\frac{u_k}{\nu^k}}\right),
\nn\\
K_u&=&-\nu\sqrt{1+z^2}\,K_\nu(\nu z)\left(\frac{u}{\nu\sqrt{1+z^2}}
+\frac{\sum\limits_{k\ge0}(-1)^k\frac{v_k}{\nu^k}}{\sum\limits_{k\ge0}(-1)^k
\frac{u_k}{\nu^k}}\right),
\label{IK}\eea
where $u_k$ and $v_k$ are the known Debye polynomials. Now, to leading order for large $\nu$, we get for the expressions in the brackets
\bea
I_u&=&\nu\sqrt{1+z^2}\,I_\nu(\nu z)\left(
1+\frac{v_1-u_1}{\nu}+\frac{u}{\nu\sqrt{1+z^2}}+O\left(\frac{1}{\nu^2}\right)
\right),
\nn\\
K_u&=&-\nu\sqrt{1+z^2}\,K_\nu(\nu z)\left(
1-\frac{v_1-u_1}{\nu}+\frac{u}{\nu\sqrt{1+z^2}}+O\left(\frac{1}{\nu^2}\right)
\right).
\eea
These expressions have to be inserted into \Ref{Alls}. It is clear that the factors $\nu\sqrt{1+z^2}$ cancel and that the expansion in the brackets becomes
\be
\frac{1+\frac{v_1-u_1}{\nu}+\frac{u}{\nu\sqrt{1+z^2}}+O\left(\frac{1}{\nu^2}\right)}
{1-\frac{v_1-u_1}{\nu}+\frac{u}{\nu\sqrt{1+z^2}}+O\left(\frac{1}{\nu^2}\right)}
=1+2\frac{v_1-u_1}{\nu}+O\left(\frac{1}{\nu^2}\right).
\ee
The dependence on $u$ drops out and the remaining factor is the correction factor which must be inserted into the calculation of the Dirichlet case. We note that the sign coming from the Bessel function $K_\nu(\nu z)$ in \Ref{IK} compensates the changed sign in \Ref{EN}.
These are the only changes which come in and performing the corresponding calculation one comes to the second line in \Ref{sph}. In this way the first correction beyond PFA for Neumann boundary conditions is calculated. From the above discussion it is evident that in leading order the same result as for Dirichlet boundary conditions appears thus re-confirming again the PFA which is insensitive to the boundary condition and that the first correction beyond does not depend on the type of Neumann condition, i.e., it does not depend on the parameter $u$.

This work was supported by the research funding from the EC's Sixth Framework Programme within the STRP project "PARNASS" (NMP4-CT-2005-01707).\\
V.N. was supported by the Swedish Research Council (Vetenskapsr{\aa}det), grant 621-2006-3046.

\section*{Appendix}
In this appendix we calculate the asymptotic expansion of the matrix elements $A_{l,l'}$, \Ref{Alls}. We start with the substitution  $l''=l+l'-2\nu$ of the summation index,
\bea\label{Alls1}A_{l,l'}&=&\sqrt{\frac{\pi}{2}}\sqrt{\frac{(2l+1)(2l'+1)}{2a\om}}
(-1)^{l+l'}\frac{I_{l'+\frac12}(\om R)}{K_{l+\frac12}(\om R)}
 \\ &&\nn
\cdot\sum_{\nu=0}^{l+l'-|l-l'|} (-i)^{l''}(2l''+1)K_{l''+\frac12}(2a\om)
\left(\begin{array}{ccc}l''&l&l'\\0&0&0\end{array}\right)
\left(\begin{array}{ccc}l''&l&l'\\0&-m&m\end{array}\right).
\eea
We are interested in the asymptotic expansion of $A_{l+l_i,l+l_{i+1}}$ for $\ep\to0$ with the variables substituted according to \Ref{subst}. To obtain the asymptotic expansion of $A_{l+l_i,l+l_{i+1}}$ we insert the asymptotic expansions for the Bessel functions and for the 3j-symbols and extend the sum over  $\nu$ till infinity. We face with the necessity to derive the asymptotic expansion of the 3j-symbols, whereas the asymptotic expansion for the Bessel functions is well known.

In general, asymptotic expansions of the 3j-symbols, or of the related by means of
\be
C^{j_3,m_3}_{j_1,m_1;j_2,m_2}=(-1)^{j_1-j_2+j_3}\sqrt{2j_3+1}
\left(\begin{array}{ccc}j_1&j_2&j_3\\m_1&m_2&-m_3\end{array}\right)
\label{CG3j}\ee
Clebsch-Gordan coefficients, are  well investigated. However, the main attentions is paid usually to  semiclassical expansions aiming for computational tools for large quantum numbers. A comprehensive and modern representation is given in \cite{Reinsch1999}. Unfortunately, it does not cover our case. The point is that   \cite{Reinsch1999} considers the limit where all quantum numbers become large whereas  in our case the magnetic quantum number $m$ grows slower that the other ones. Nevertheless \cite{Reinsch1999} gives a very convenient integral representation which serves as the starting point in our case too. We used Eq.(2.11) in \cite{Reinsch1999} which in our notations reads
\bea
C^{l'',0}_{l,m;l',-m}&=&\frac{(-1)^{l''}}{\pi^2}(-4)^{(l''+l+l')/2} N_{l,m;l',-m;l'',0}
\nn \\ &&\cdot
\int_{-\pi/2}^{\pi/2}\int_{-\pi/2}^{\pi/2}
e^{2im(\phi-\theta)}(\cos\phi)^{l''+l-l'}(\cos\theta)^{l''-l+l'}
\nn \\ && \cdot
(\sin(\theta-\phi))^{-l''+l+l'}
d\theta d\phi \, ,
\label{CG1}\eea
whereby we shifted the angles $\theta$ and $\phi$ by $\pi/2$ which resulted in the factors $\cos\theta$ and $\cos\phi$ in place to the factors $\sin\theta$ and $\sin\phi$ in \cite{Reinsch1999} . The factor $N_{l,m;l',-m;l'',0}$ is defined by
\bea\label{N1}
N_{l,m;l',-m;l'',0}&=&\sqrt{(4l+2n_i+2n_{i+1}-4\nu+1)}\nn \\ && \times
\left(\frac{(l+n_i+m)!(l+n_i-m)!(l+n_{i+1}+m)!(l+n_{i+1}-m)! }{(4l+2n_i+2n_{i+1}-2\nu+1)!(2\nu)!}
\right)^{\frac12}\nn \\ &&\times
\frac{ (2l+n_i+n_{i+1}-2\nu)!}{\left((2l+2n_i-2\nu)!(2l+2n_{i+1}-2\nu)!
\right)^\frac12}
\eea
in our notations.

The asymptotic region we are interested in is determined by Eq.\Ref{subst}. So we have to use $l''=l+l'-2\nu$, $\l\to l+l_i$ and $l'\to l+l_{i+1}$ in Eqs.\Ref{CG1} and \Ref{N1} to perform calculations in the asymptotic region. First of all we consider $N_{l,m;l',-m;l'',0}$. Its asymptotic expansion follows simply from  Stirlings formula. For small $\ep$ it holds
\bea \left(\frac{a}{\ep}+\frac{b}{\sqrt{\ep}}+c\right)!&=&
\frac{\sqrt{2\pi a}}{\ep}
\exp\left[
\frac{a}{\ep}\left(-1+\ln\frac{a}{\ep}\right)+\frac{b}{\sqrt{\ep}}\ln\frac{a}{\ep}
+c \ln\frac{a}{\ep}+\frac{b^2}{2a}
\right]
\nn \\ && \cdot
\Big(1+\frac{b \left(a (6 c+3)-b^2\right) \sqrt{\ep}}{6
   a^2}
\nn \\ &&
   +\frac{\left(b^6-12 a c b^4+9 a^2 \left(4 c^2-1\right) b^2+6 a^3
   \left(6 c^2+6 c+1\right)\right) \ep}{72
   a^4}
\nn \\ &&
    +O\left(\ep^{3/2}\right)
   \Big).
\label{N2}\eea
Now we make the substitutions \Ref{subst} in $N_{l,m;l',-m;l'',0}$, \Ref{N1}, and obtain with  \Ref{N2}
\be
N_{l,m;l',-m;l'',0}=
\frac{\pi^{3/4}}{\sqrt{(2\nu)!}}\, 2^{-4l-2(l_i+l_{i+1})+2\nu+1/4}\, l^{\nu+3/4} \, e^{m^2/l}  \, \tilde{N}(\ep)\label{N3}\ee
where the function
\bea \tilde{N}(\ep)&=&
 1-\frac{\left(({n_i}+{n_{i+1}}) \left(4 \mu ^2-4 \nu -3\right)\right)
  }{4 \left(\sqrt{t} \tau \right)} \sqrt{{\ep}}
  \nn \\ &&
   +\frac{1}{{96 t \tau^2}}
    \left(3 \left(16 \mu
   ^4-8 (4 \nu -5) \mu ^2+16 \nu ^2-24 \nu -11\right) {(n_i+n_{i+1})}^2
\right. \nn \\ && \left.
+96(1-4\mu^2+4\nu) n_in_{i+1}   
\right.\nn \\ &&\left.
    +\left(16 \mu ^4-48 \mu ^2-3 \left(8 \nu ^2+4 \nu
   -5\right)\right) \tau \right) {\ep}
\nn  \\  &&
    +O\left({\ep}^{3/2}\right)
\label{N3a}\eea
collects the subleading contributions.

Next we calculate the integral in \Ref{CG1} using the saddle point method. The considered integral $I$ we rewrite  in the form
\bea        I=
\int_{-\pi/2}^{\pi/2}\int_{-\pi/2}^{\pi/2}  (\theta-\phi)^{2\nu}
e^{2im (\theta-\phi)}
g(\theta,\phi) \, e^{-(l''+l-l')h(\phi)-(l''-l+l')h(\theta)}
d\theta d\phi
\label{CG2}\eea
with
\be
h(\theta)= \ln {\cos\theta},
\qquad
g(\theta,\phi)=\left(\frac{\sin(\theta-\phi)}{\theta-\phi}\right)^{2\nu}\, \, .
\label{hg}\ee
For small $\ep$, because of $l''+l-l'=2t\tau/\ep+\sqrt{4t/\ep}(n_i-n_{i+1})-2\nu$, the dominating contributions come  from the minimum of the functions $h$ which are reached in $\theta=0$ and $\phi=0$. Therefore we expand these functions using $h(x)=x^2/2+x^4/12+\dots$.  The leading order contribution in the exponential is then $\exp(-\frac{2t\tau}{\ep}(\theta^2+\phi^2))$. Taking into account that the function $g$ depends on the difference, $\theta-\phi$, we are motivated to make the substitution of variables
\be
\theta =\sqrt{\frac{\ep}{2t\tau}} \ (\xi+\eta), \qquad
\phi   =\sqrt{\frac{\ep}{2t\tau}} \ (\xi-\eta) \, .
\label{subst2}\ee
After that the integral can be rewritten now in the form
\bea        I= \left(\frac{2\ep}{t\tau}\right)^{1+\nu}
\int_{-\infty}^{\infty}\int_{-\infty}^{\infty}  y^{2\nu}\, e^{2i\sqrt{2\tau}\mu\eta}\,
\tilde{g}(\xi,\eta) \, e^{-(\xi^2+\eta^2)}
d\xi\, d\eta \, ,
\label{CG3}\eea
In the sense of the asymptotic expansion we are interested in we stretched the integration until infinity. In \Ref{CG3} we defined the function
\be
\tilde{g}(\xi,\eta)=g(\theta,\phi) \, e^{-(l''+l-l')h(\phi)-(l''-l+l')h(\theta)+\xi^2+\eta^2}.
\label{gtilde}\ee
Obviously, it has an expansion in powers of $\sqrt{\ep}$,
\bea
\tilde{g}(\xi,\eta)&=&
1+
    \left(-({n_i}+{n_{i+1}}) \eta ^2+2
   ({n_{i+1}}-{n_i}) \xi  \eta
   -({n_i}+{n_{i+1}}) \xi ^2\right)
  \frac{\sqrt{{\ep}}}{\sqrt{t} \tau   }
\nn \\ &&
+
   \left(
   \left(6 {(n_i+n_{i+1})}^2-\tau \right) \eta ^4+24
   \left({n_i}^2-{n_{i+1}}^2\right) \xi  \eta^3
\right.\nn \\ && \left.
    +\left(6 \left(6 {n_i}^2-4 {n_{i+1}}
   {n_i}+6 {n_{i+1}}^2-\tau \right) \xi ^2+4 \nu
   \tau \right) \eta ^2
\right.\nn \\ &&\left.
   +24
   \left({n_i}^2-{n_{i+1}}^2\right) \xi ^3 \eta
\right.\nn \\ && \left.
   +\left(6 {(n_i+n_{i+1})}^2-\tau \right) \xi ^4+12 \nu  \tau  \xi^2
  \right) \frac{{\ep}}{12 t \tau   ^2}
   +O\left({\ep}^{3/2} \right) \, .
\label{gtilde1}
\eea
Now we carry out the integration over $\xi$. It is Gaussian and delivers a new function $\tilde{g}_1(\ep,\eta)$,
\bea        I=\sqrt{\pi}\left(\frac{2\ep}{t\tau}\right)^{1+\nu} \,
\int_{-\infty}^{\infty}  y^{2\nu}
\tilde{g}_1(\ep,\eta) \, e^{-\eta^2+2i\sqrt{2\tau}\mu\eta}
\, d\eta \,
\label{CG4}\eea
with the property $\tilde{g}_1(0,\eta)=1$.
Finally, we take the expansion \Ref{N3} of the prefactor and the expansion \Ref{CG4} of the integral together and insert  into Eq.\Ref{CG1} for the Clebsch-Gordan coefficients. This gives
\bea
C^{l'',0}_{l,m;l',-m}&=&
\frac{(-1)^\nu}{\pi^{1/4}}\, 2^{-\nu+1/4} \frac{\sqrt{(2\nu)!}}{\nu!}\, l^{-1/4} h(\ep,\mu)\, ,
\label{CG5}\eea
where    the new function $h(\ep,\mu)$ collects the sub-leading contributions,
\be
h(\ep,\mu)=\frac{2^{2\nu}\nu!}{\sqrt{\pi}(2\nu)!}\int_{-\infty}^\infty \eta^{2\nu}g_2(\ep,\eta) \, e^{2i\sqrt{2}\mu\eta-\eta^2+\mu^2}d\eta\,.
\label{CG6}\ee
Here we defined $g_2(\ep,\eta)=n_1(\ep)\tilde{g}_1(\ep,\eta)$ which must be re-expanded in powers of $\sqrt{\ep}$ too. The function $h(\ep,\mu)$ has the property $h(0,\mu)=1$. It is possible to obtain a more explicit representation of this function by carrying out the integration. However, for the application below this proved to be not useful.

The calculation of the asymptotic expansion \Ref{CG5} of the Clebsch-Gordan coefficients was the most difficult task. It remains to insert them into the matrix elements $A_{l,l'}$, \Ref{Alls1}. These symbols enter twice. The second time they go with zero magnetic quantum number $m$. It is possible to put $m=0$ directly in the asymptotic formula \Ref{CG5}. In \Ref{CG6} then the integration can be carried out explicitly.
Putting these formulas together with the well known asymptotic expansion of the Bessel functions (which is displayed, for instance, in Eq.(B.1) in  \cite{Bordag:2006vc}) and we come to
\be
A_{l+l_i,l+l_{i+1}} =
\frac{\tau e^{-\eta_{as}}}{\sqrt{2\pi l (1-\tau^2)}}
\int_{-\infty}^\infty\frac{d\eta}{\sqrt{\pi}}\
e^{-\eta^2+2i\sqrt{2}\mu\eta+\mu^2}
\sum_{\nu=0}^\infty\frac{\eta^{2\nu}}{\nu!}
\left(\frac{1-\tau}{1+\tau}\right)^\nu C_g
\label{Alls2}\ee
where $C_g$ collects the sub-leading factors. Its dependence on $\nu$ is polynomial. Hence, the sum over $\nu$ can be carried out,
\be
\sum_{\nu=0}^\infty\frac{\eta^{2\nu}}{\nu!}
\left(\frac{1-\tau}{1+\tau}\right)^\nu C_g=
e^{\eta^2\frac{1-\tau}{1+\tau}}C_a,
\label{sumnu}\ee
where now $C_a$ collects the non-leading contributions. The complete expression is too lengthy to be shown here. So we restrict ourselves to show the order $\sqrt{\ep}$,
\bea
C_a&=&1+\frac{1}{2   \sqrt{t} \tau  (\tau +1)}
   \Big(
   2 \tau ^2 (\tau +1) n_i^3-2 n_{i+1} \tau
   ^2 (\tau +1) n_i^2
\nn \\ &&
    +\left(\left(-2
   n_{i+1}^2-4 t+1\right) \tau ^3
    +\left(4 \eta ^2-2
   n_{i+1}^2-4 t\right) \tau ^2
   -8 \eta ^2 \tau
\right.\nn \\ && \left.
   +\tau
   -2 \mu ^2 (\tau +1)+2\right) n_i+n_{i+1}
   \left(\left(2 n_{i+1}^2-4 t-3\right) \tau ^3
\right. \\ \nn&& \left.
   +2
   \left(2 \eta ^2+n_{i+1}^2-2 t-2\right) \tau ^2-8
   \eta ^2 \tau +\tau -2 \mu ^2 (\tau +1)+2\right)
   \Big) \sqrt{\ep}.
\label{Ca}\eea
With \Ref{sumnu} the integration over $\eta$ in \Ref{Alls2} becomes Gaussian and can be carried out. Together with the remaining factors the result collects into the final form of the matrix elements up to the $\ep$ order,
\be
 A_{l+l_i,l+l_{i+1}}=\sqrt{\frac{\ep}{4\pi t}}
\, e^{-\eta_{as}}\, e^{-\mu^2}
\left(
1+a^{\frac12}(n_i,n_{i+1})\sqrt{\ep}
+a^{1}(n_i,n_{i+1})\ep \right),\nn \\
\label{Alls3}\ee
where
\be
\eta_{as}=-2t+(n_i-n_{i+1})^2
\label{etaas}\ee
is the same factor as in the cylindrical case  \cite{Bordag:2006vc} and
the non-leading coefficients are
\bea
a^{\frac12}_{n_i,n_{i+1}}&=&\frac{1}{2 \sqrt{t} \tau ^2}
\Big( 2 n_{i}^3 \tau ^3-2 n_{i}^2 n_{i+1}
   \tau ^3
   \nn \\ &&
   +n_{i} \left(\left(-2 n_{i+1}^2-4
   t+1\right) \tau ^3-2 \mu ^2 \left(\tau
   ^2-2\right)\right)
\nn \\ &&
   +n_{i+1} \left(\left(2
   n_{i+1}^2-4 t-3\right) \tau ^3-2 \mu ^2 \left(\tau
   ^2-2\right)\right)   \Big)\ \label{}
      \\
a^{1}_{n_i,n_{i+1}}&=&   \frac{1}{48 t
   \tau ^4}     \Big(
   4 \left(3 \tau ^4-9 \tau ^2+6 n_{i}^2
   \left(\tau ^2-2\right)^2+6 n_{i+1}^2 \left(\tau
   ^2-2\right)^2
\nn \right. \\ &&\left.
   +12 n_{i} n_{i+1} \left(\tau
   ^2-2\right)^2+4\right) \mu ^4-12 \tau  \left(4 \tau
   ^2 \left(\tau ^2-2\right) n_{i}^4
\nn \right. \\ &&\left.
   -2 \left(4
   \left(n_{i+1}^2+t\right) \tau ^4+\left(-8
   n_{i+1}^2-8 t+5\right) \tau ^2-8\right)
   n_{i}^2
\nn \right. \\ &&\left.
   -4 n_{i+1} \left((4 t+2) \tau ^4+(1-8
   t) \tau ^2-4\right) n_{i}-4 t \tau ^4-\tau ^4+2
   \tau ^3
\nn \right. \\ &&\left.
   +4 t \tau ^2-3 \tau ^2-4 \tau +4 n_{i+1}^4
   \tau ^2 \left(\tau ^2-2\right)
\nn \right. \\ &&\left.
   +n_{i+1}^2 \left(-8
   (t+1) \tau ^4+2 (8 t+3) \tau ^2+16\right)+4\right)
   \mu ^2
\nn  \\ &&
   +\tau ^2 \left(3 \left(8 n_{i}^6-16
   n_{i+1} n_{i}^5-4 \left(2 n_{i+1}^2+8
   t+5\right) n_{i}^4
\right.\right.\nn  \\ &&  \left.\left.
   +16 n_{i+1} \left(2
   n_{i+1}^2-1\right) n_{i}^3+\left(-8
   n_{i+1}^4+8 (8 t+5) n_{i+1}^2
\right.\right.\right.\nn  \\ &&  \left.\left.   \left.
   +32 t^2-22\right)
   n_{i}^2
\right.\right.\nn  \\ &&  \left.\left.
   +4 n_{i+1} \left(-4 n_{i+1}^4+12
   n_{i+1}^2+16 t^2+16 t+1\right) n_{i}
 \right.\right.\nn  \\ &&  \left.\left.
   +8
   n_{i+1}^6+16 t^2+8 t-4 n_{i+1}^4 (8
   t+13)
\right.\right.\nn  \\ &&  \left.\left.
   +n_{i+1}^2 \left(32 t^2+64
   t+58\right)-5\right) \tau ^4+12 \left(2
   n_{i}^2-4 n_{i+1} n_{i}+2 n_{i+1}^2
\right.\right.\nn  \\ &&  \left.\left.
   -4
   t-1\right) \tau ^3+\left(28 n_{i}^4-16 n_{i+1}
   n_{i}^3-12 \left(2 n_{i+1}^2+4 t-3\right)
   n_{i}^2
\right.\right.\nn  \\ &&  \left.\left.
   -8 n_{i+1} \left(2 n_{i+1}^2+12
   t+3\right) n_{i}+28 n_{i+1}^4-24 t
\right.\right.\nn  \\ &&  \left.\left.
   -12
   n_{i+1}^2 (4 t+5)+9\right) \tau ^2+12\right) \Big).\ \label{} \eea

\end{document}